\documentclass[11pt]{article}
\usepackage{moriond,epsfig}

\def\be{\begin{equation}}
\def\ee{\end{equation}}
\def\bea{\begin{eqnarray}}
\def\eea{\end{eqnarray}}

\def\dd{\mathrm{d}}

\def\jpsi{J/\psi}
\def\psip{\psi^\prime}

\def\ee{e^+ e^-}

\def\fz{\footnotesize}
\def\sz{\scriptsize}
\begin{document}
\vspace*{4cm}
\title{New results on nuclear dependence \\of $\jpsi$ and $\psip$ production
in 450 GeV pA collisions}

\author{ R. SHAHOYAN for NA50 Collaboration\\
M.C.~Abreu$^{6,a}$,
B.~Alessandro$^{10}$,
C.~Alexa$^{3}$,
R.~Arnaldi$^{10}$,
M.~Atayan$^{12}$,
C.~Baglin$^{1}$,
A.~Baldit$^{2}$,
M.~Bedjidian$^{11}$,
S.~Beol\`e$^{10}$,
V.~Boldea$^{3}$,
P.~Bordalo$^{6,b}$,
S.R.Borenstein$^{9,c}$,
G.~Borges$^{6}$,
A.~Bussi\`ere$^{1}$,
L.~Capelli$^{11}$,
C.~Castanier$^{2}$,
J.~Castor$^{2}$,
B.~Chaurand$^{9}$,
B.~Cheynis$^{11}$,
E.~Chiavassa$^{10}$,
C.~Cical\`o$^{4}$,
T.~Claudino$^{6}$,
M.P.~Comets$^{8}$,
N.~Constans$^{9}$,
S.~Constantinescu$^{3}$,
P.~Cortese$^{10,d}$,
J.~Cruz,$^{6}$,
N.~De Marco$^{10}$,
A.~De Falco$^{4}$,
G.~Dellacasa$^{10,d}$,
A.~Devaux$^{2}$,
S.~Dita$^{3}$,
O.~Drapier$^{11}$,
B.~Espagnon$^{2}$,
J.~Fargeix$^{2}$,
P.~Force$^{2}$,
M.~Gallio$^{10}$,
Y.K.~Gavrilov$^{7}$,
C.~Gerschel$^{8}$,
P.~Giubellino$^{10}$,
M.B.~Golubeva$^{7}$,
M.~Gonin$^{9}$,
A.A.~Grigorian$^{12}$,
S.~Grigorian$^{12}$,
J.Y.~Grossiord$^{11}$,
F.F.~Guber$^{7}$,
A.~Guichard$^{11}$,
H.~Gulkanyan$^{12}$,
R.~Hakobyan$^{12}$,
R.~Haroutunian$^{11}$,
M.~Idzik$^{10,e}$,
D.~Jouan$^{8}$,
T.L.~Karavitcheva$^{7}$,
L.~Kluberg$^{9}$,
A.B.~Kurepin$^{7}$,
Y.~Le~Bornec$^{8}$,
C.~Louren\c co$^{5}$,
P.~Macciotta$^{4}$,
M.~Mac~Cormick$^{8}$,
A.~Marzari-Chiesa$^{10}$,
M.~Masera$^{10}$,
A.~Masoni$^{4}$,
M.~Monteno$^{10}$,
A.~Musso$^{10}$,
P.~Petiau$^{9}$,
A.~Piccotti$^{10}$,
J.R.~Pizzi$^{11}$,
W.~Prado da Silva$^{10,f}$,
F.~Prino$^{10}$,
G.~Puddu$^{4}$,
C.~Quintans$^{6}$,
S.~Ramos$^{6,b}$,
L.~Ramello$^{10,d}$,
P.~Rato Mendes$^{6}$,
L.~Riccati$^{10}$,
A.~Romana$^{9}$,
H.~Santos$^{6}$,
P.~Saturnini$^{2}$,
E.~Scalas$^{10,d}$,
E.~Scomparin$^{10}$
S.~Serci$^{4}$,
R.~Shahoyan$^{6,g}$,
F.~Sigaudo$^{10}$,
S.~Silva$^{6}$,
M.~Sitta$^{10,d}$,
P.~Sonderegger$^{5,b}$,
X.~Tarrago$^{8}$,
N.S.~Topilskaya$^{7}$,
G.L.~Usai$^{4}$,
E.~Vercellin$^{10}$,
L.~Villatte$^{8}$,
N.~Willis$^{8}$.\\
{\sz $^{~1}$ LAPP, CNRS-IN2P3, Annecy-le-Vieux,  France.
$^{~2}$ LPC, Univ. Blaise Pascal and CNRS-IN2P3, Aubi\`ere, France.
$^{~3}$ IFA, Bucharest, Romania.
$^{~4}$ Universit\`a di Cagliari/INFN, Cagliari, Italy.
$^{~5}$ CERN, Geneva, Switzerland.
$^{~6}$ LIP, Lisbon, Portugal.
$^{~7}$ INR, Moscow, Russia.
$^{~8}$ IPN, Univ. de Paris-Sud and CNRS-IN2P3, Orsay, France.
$^{~9}$ LPNHE, Ecole Polytechnique and CNRS-IN2P3, Palaiseau, France.
$^{10}$ Universit\`a di Torino/INFN, Torino, Italy.
$^{11}$ IPN, Univ. Claude Bernard Lyon-I and CNRS-IN2P3, Villeurbanne,
France.
$^{12}$ YerPhI, Yerevan, Armenia.\\
a) also at UCEH, Universidade de Algarve, Faro, Portugal
b) also at IST, Universidade T\'ecnica de Lisboa, Lisbon, Portugal
c) on leave of absence from York College CUNY
d) Universit\'a del Piemonte Orientale, Alessandria and INFN-Torino,
Italy
e) also at Faculty of Physics and Nuclear Techniques, University of
   Mining and Metallurgy, Cracow, Poland
f) now at UERJ, Rio de Janeiro, Brazil
g) on leave of absence from YerPhI, Yerevan, Armenia
}}
\address{}

\maketitle\abstracts{
To understand the reliability of the charmonia suppression as
a signature of the Quark-Gluon Plasma formation in
nucleus-nucleus collisions it is important first 
to understand the details of the production of $\jpsi$ and $\psip$
in $pA$ interactions and the difference in the suppression of these
two states. This report presents the results of the study by the NA50 
collaboration of the $\jpsi$ and $\psip$ production in $pA$ interactions 
at $450$ GeV beam energy and its dependence on rapidity.
It is shown that the $\psip$ suffers more suppression than the $\jpsi$, 
which is consistent with a similar observation made
at $800$ GeV beam energy by the E866/NuSea collaboration. 
}

\section{Introduction}
\label{sec:intro}
The suppression of the production of charmonia states (mainly $\jpsi$ and $\psip$) in 
$nucleus-nucleus$ interactions due to colour screening is recognized as one of the most 
promising QGP signatures~\cite{Mat86}. But in real experiments it should be distinguished from the 
suppression due to the hadronic interactions of charmonia with surrounding nuclear matter. This kind
of suppression, present already in pA interactions, can describe all currently available data except the
``anomalous $\jpsi$ suppression'' observed by the NA50 experiment in PbPb collisions~\cite{THEPSI}. 
The deconvolution of these two kinds of suppression requires precise quantitative understanding of the 
interaction of different charmonium states with hadronic matter. 
While a vast amount of data was collected for the $\jpsi$, the situation with the $\psip$ is less clear. Published 
NA50 data showed similar levels of suppression for both mesons in pA collisions but a stronger suppression of the
$\psip$ in heavier colliding systems, like SU~\cite{Rame98}, which was attributed to additional absorption of the
$\psip$ due to interactions with ``comovers'' - hadrons produced in the collision. Finally, the E866/NuSea 
collaboration~\cite{Lei00} observed a stronger suppression for the $\psip$ already in 800 GeV pA interactions.
The motivation of the present study is to look for similar effects in the NA50 data.

\section{Experimental setup, data samples and analysis}
\label{sec:expna50}
The main component of the NA50 apparatus is a dimuon spectrometer, covering the $3.0<y_{lab}<4.0$ 
rapidity range, which is separated from the target region by a 4.8 m long hadron absorber.
The dimuon trigger is provided by four scintillator hodoscopes and its efficiency is controlled
during special runs by two additional hodoscopes.
The detailed description of the experiment can be found in \cite{na50app}. 
We present here the analysis of data collected in the 1996-2000 period with a 450~GeV proton beam 
interacting with Be, 
Al, Cu, Ag and W targets. Each data sample was obtained with single target, with interaction
probability varying from 26 (Al) to 39 (Ag)$\%$. The incoming beam intensity was monitored by three 
argon ionization chambers. For each target two qualitatively different data samples were collected:
with ``low'' ({\bf LI}, $1-3\times 10^{8}$ protons per $2.37$ s spill) and ``high'' 
({\bf HI}, $1-3\times 10^{9}$ p/spill) beam intensities. While the {\bf LI} samples
\footnote{The results for the {\bf LI} sample were presented in \cite{EnricoQM} 
(obtained with a somewhat different analysis procedure).}
are more reliable from the absolute normalization point of view, 
the {\bf HI} one provides much better statistics, which is necessary for the
$\psip$ study. 

The reconstructed dimuons were subjected to $-0.5 <y_{cm}<0.5$ 
and $|\cos{\theta_{CS}}|<0.5$ selection cuts. For the differential cross section studies, the rapidity range was 
divided in four equidistant bins. The mass spectra in each kinematical bin were fitted with 
curves corresponding to individual dimuon sources, obtained from detailed Monte-Carlo simulations.

Combinatorial background from uncorrelated $\pi,K \rightarrow \mu$ decays
was estimated from the like-sign dimuon samples (using the relation 
$N_{+-} = 2 R \sqrt{N_{++} N_{--}}$, with $R$ being a free parameter of the fit, accounting for possible
charge correlations in parent hadrons production). The contribution from the dimuons originated in
interactions outside of the target was accounted by including into the fit the dimuon spectrum obtained
from the special ``target-out'' runs.

The systematic errors of the extracted cross sections account for uncertainties in the normalization:
luminosity, trigger and reconstruction efficiencies.

Since various authors use different parametrizations for the description of charmonia suppression, we 
fitted the obtained cross sections by three commonly used models:\\
1) The Glauber model, assuming that each charmonium state $i$ is produced in binary 
nucleon-nucleon interactions with cross section $\sigma_{0}^{i}$ and then interacts with 
surrounding nuclear matter with cross section
$\sigma_{abs}^{i}$. This leads to the pA cross section 
\( \sigma_{pA} = \sigma_{o}/\sigma_{abs}^2 \int \dd \vec{s} ~ {\left[ {1 - T_{A}(\vec{s})} 
\sigma_{abs} \right] }^{A} \label{eq_sigpsiab1} \), where $T_{A}(\vec{s})$ is the nuclear thickness 
function at impact vector $\vec{s}$.\\
2) A simplification of the Glauber model, assuming that the charmonium is absorbed with cross section 
$\sigma_{0}$ seeing in average $<\rho L>$ amount of matter from its production point to its exit from the nucleus:
\(\sigma_{pA} = \sigma_{o} ~A e^{-\sigma_{abs} <\rho L>}\)  
($<\rho L>= (A-1)/2  \int \dd \vec{s} ~ T_{A}^2(\vec{s})$ is obtained from the expansion of the Glauber formula).\\
3) The widely used although not theory-motivated parametrization $\sigma_{pA} = \sigma_{o} A^{\alpha}$.

\section{Results}
\label{sec:results} 

Table \ref{tab_y_dst_psi} shows the $\jpsi$ and $\psip$ cross sections in 
the whole rapidity range (first column) and in four separate bins (columns 2-5). 
Due to the lack of space we show the cross sections obtained from the {\bf HI} data sample only. 
Although the cross sections 
extracted from the {\bf LI} data sample are systematically higher, it is the same $\sim 5\%$ difference both
for the $\jpsi$ and the $\psip$, which justifies the treatment of this discrepancy as a constant normalization factor.
One should notice that since the
systematic errors reflect the uncertainty in the absolute normalizations, they do not affect the shapes of the
differential distributions in rapidity.
\vspace{-5mm}
\begin{table}[ht]
\caption{$\jpsi$ and $\psip$ cross sections per nucleon in $nb$ (not corrected for the $\mu \mu$ branching ratio) 
for the integrated data and in four $Y_{CM}$ bins. The statistical and systematic errors are shown 
separately.\label{tab_y_dst_psi}}
\begin{center}
\begin{tabular}{|l|c||c|c|c|c|}
\hline
&\fz \bf -0.50$<Y<$0.50&\fz \bf -0.50$<Y<$-0.25&\fz \bf -0.25$<Y<$ 0.0&\fz \bf 0.0$<Y<$ 0.25&\fz \bf 0.25$<Y<$ 0.5\\ \hline
\multicolumn{6}{|c|}{$\jpsi$} \\ \hline
 \small \bf pBe &\sz 5.130$\pm$0.010$\pm$0.177 &\sz 1.202$\pm$0.007$\pm$0.042 &\sz 1.373$\pm$0.004$\pm$0.047 &\sz 1.352$\pm$0.004$\pm$0.047 &\sz 1.210$\pm$0.007$\pm$0.042\\
 \small \bf pAl &\sz 4.868$\pm$0.008$\pm$0.228 &\sz 1.117$\pm$0.005$\pm$0.053 &\sz 1.304$\pm$0.003$\pm$0.061 &\sz 1.281$\pm$0.003$\pm$0.060 &\sz 1.145$\pm$0.005$\pm$0.054\\
 \small \bf pCu &\sz 4.712$\pm$0.006$\pm$0.181 &\sz 1.173$\pm$0.004$\pm$0.045 &\sz 1.275$\pm$0.003$\pm$0.049 &\sz 1.209$\pm$0.003$\pm$0.047 &\sz 1.069$\pm$0.004$\pm$0.041\\
 \small \bf pAg &\sz 4.403$\pm$0.005$\pm$0.148 &\sz 1.077$\pm$0.004$\pm$0.036 &\sz 1.190$\pm$0.002$\pm$0.040 &\sz 1.134$\pm$0.002$\pm$0.038 &\sz 1.016$\pm$0.004$\pm$0.034\\
 \small \bf pW &\sz 4.005$\pm$0.006$\pm$0.147 &\sz 0.945$\pm$0.004$\pm$0.035 &\sz 1.068$\pm$0.003$\pm$0.039 &\sz 1.047$\pm$0.003$\pm$0.038 &\sz 0.945$\pm$0.004$\pm$0.035\\ \hline
\multicolumn{6}{|c|}{$\psip$} \\ \hline
\small \bf pBe &\sz .0886$\pm$.0021$\pm$.0032 &\sz .0207$\pm$.0014$\pm$.0008 &\sz .0231$\pm$.0010$\pm$.0008 &\sz .0228$\pm$.0009$\pm$.0008 &\sz .0217$\pm$.0013$\pm$.0009\\
\small \bf pAl &\sz .0841$\pm$.0015$\pm$.0044 &\sz .0194$\pm$.0010$\pm$.0009 &\sz .0215$\pm$.0006$\pm$.0011 &\sz .0228$\pm$.0006$\pm$.0012 &\sz .0209$\pm$.0009$\pm$.0010\\
\small \bf pCu &\sz .0773$\pm$.0011$\pm$.0032 &\sz .0189$\pm$.0007$\pm$.0008 &\sz .0210$\pm$.0005$\pm$.0008 &\sz .0197$\pm$.0004$\pm$.0009 &\sz .0173$\pm$.0007$\pm$.0009\\
\small \bf pAg &\sz .0690$\pm$.0010$\pm$.0025 &\sz .0151$\pm$.0006$\pm$.0005 &\sz .0189$\pm$.0004$\pm$.0007 &\sz .0182$\pm$.0004$\pm$.0007 &\sz .0161$\pm$.0006$\pm$.0005\\
\small \bf pW &\sz .0611$\pm$.0010$\pm$.0024 &\sz .0130$\pm$.0006$\pm$.0005 &\sz .0161$\pm$.0004$\pm$.0006 &\sz .0162$\pm$.0004$\pm$.0007 &\sz .0158$\pm$.0007$\pm$.0006\\ \hline
\end{tabular}
\end{center}
\end{table}
\vspace{-4mm}

\begin{figure}[ht]
\vspace{-2mm}
\centering
\begin{tabular}{c}
\resizebox{0.33\textwidth}{!}{%
\includegraphics*{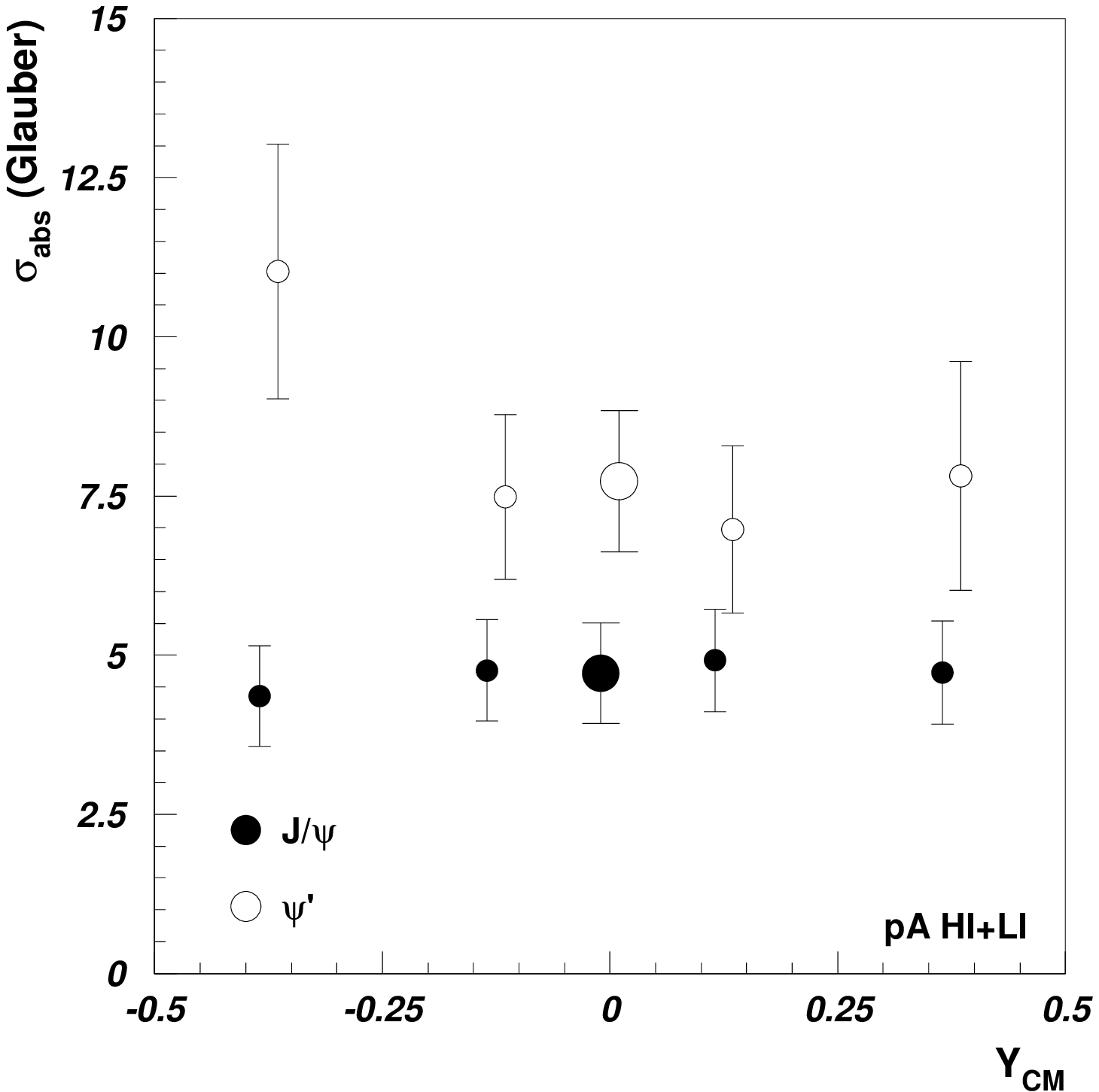}}
\resizebox{0.33\textwidth}{!}{%
\includegraphics*{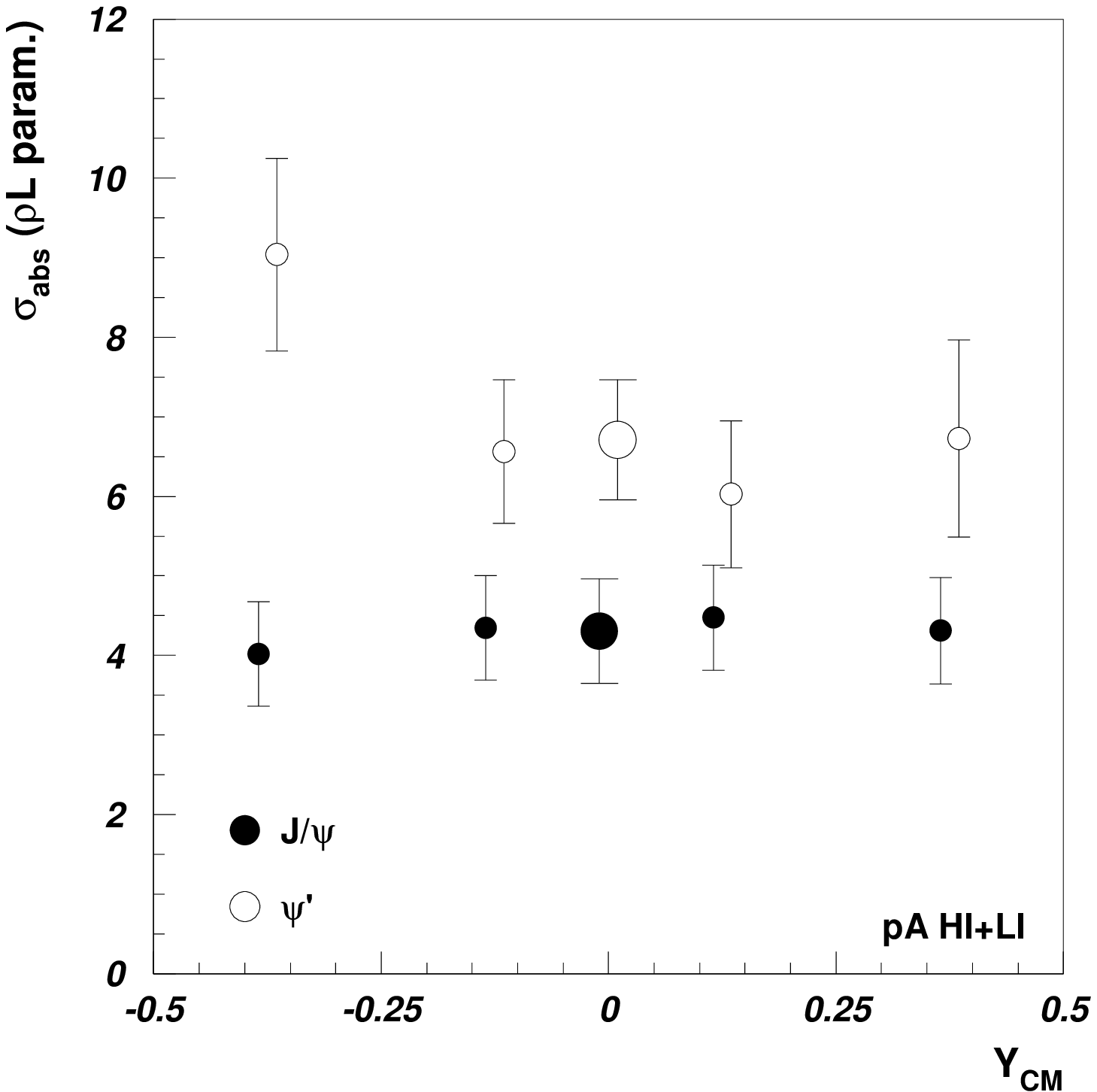}}
\resizebox{0.33\textwidth}{!}{%
\includegraphics*{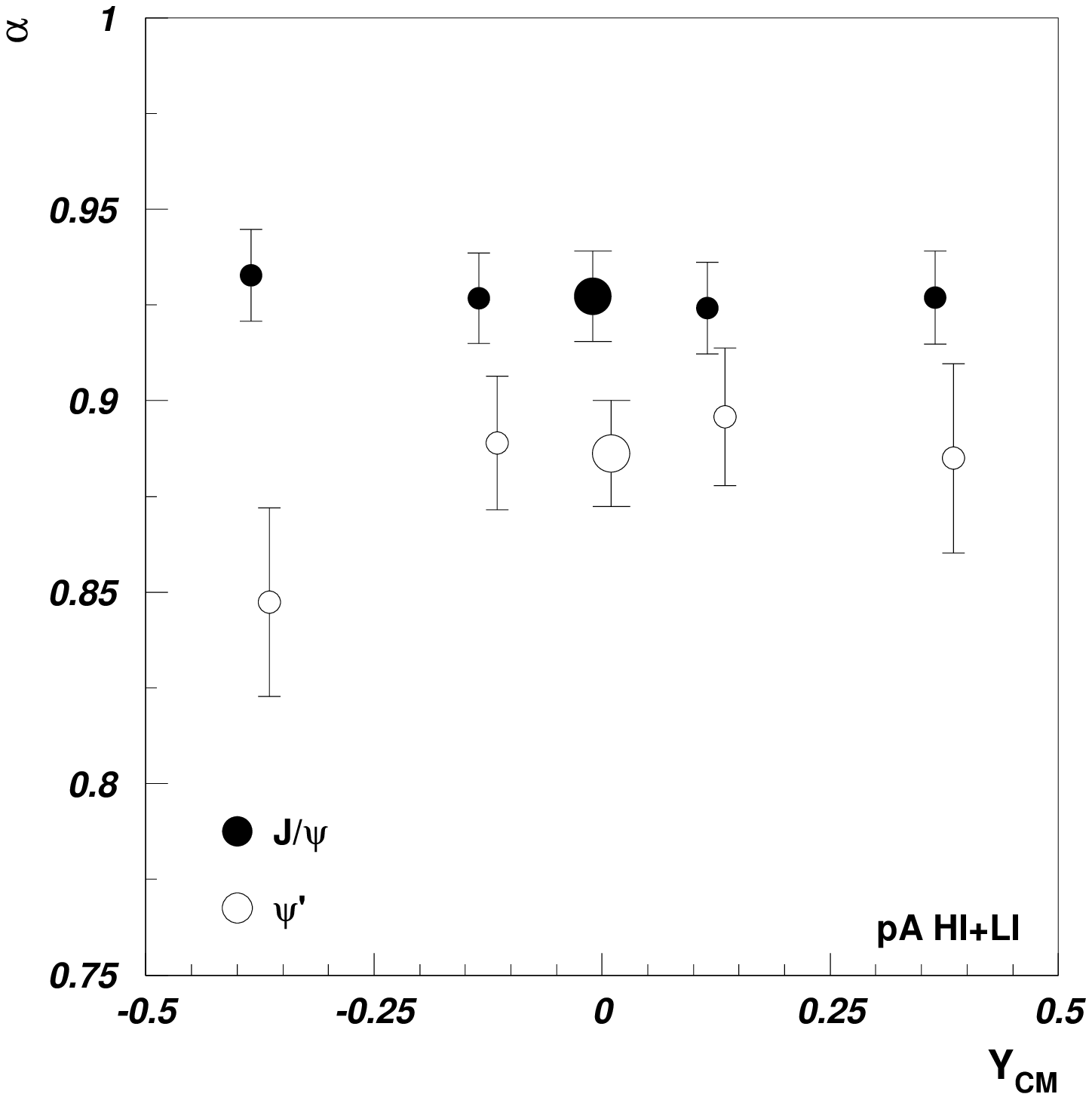}}
\end{tabular}
\vspace{-6mm}
\caption{ $\alpha$ parameters and absorption cross sections 
for the $\jpsi$ and $\psip$ as a function of the
$Y_{CM}$ obtained from the absolute cross sections fit: (Left) 
Glauber model; (Center) $e^{-<\rho L>\sigma_{abs}}$; (Right) $A^{\alpha}$ parametrizations. 
The large symbols in the center correspond to fitting the data integrated over $-0.5<y_{cm}<0.5$ .
\label{fig_alpha_hili}}
\end{figure}

Fig. \ref{fig_alpha_hili} shows the $\jpsi$ and $\psip$ absorption parameters obtained from the global fit of {\bf HI} and {\bf LI}
data samples by the models mentioned in Sec.\ref{sec:expna50}. 
For these joint fits the cross sections from the {\bf HI} and {\bf LI} 
samples were rescaled by the factors $2/(1+R)$ and $2R/(1+R)$, respectively ($R=0.957$ is the result of the fit of 
the ratio of {\bf HI} to {\bf LI}  values by a constant line). The results obtained separately from {\bf HI} 
and {\bf LI} samples are similar to the shown results. 
The large symbols in the center correspond to the fit of the data integrated over the $-0.5<y_{cm}<0.5$ range.

\begin{figure}[ht]
\centering
\vglue -2mm
\begin{tabular}{c}
\resizebox{0.33\textwidth}{!}{%
\includegraphics*{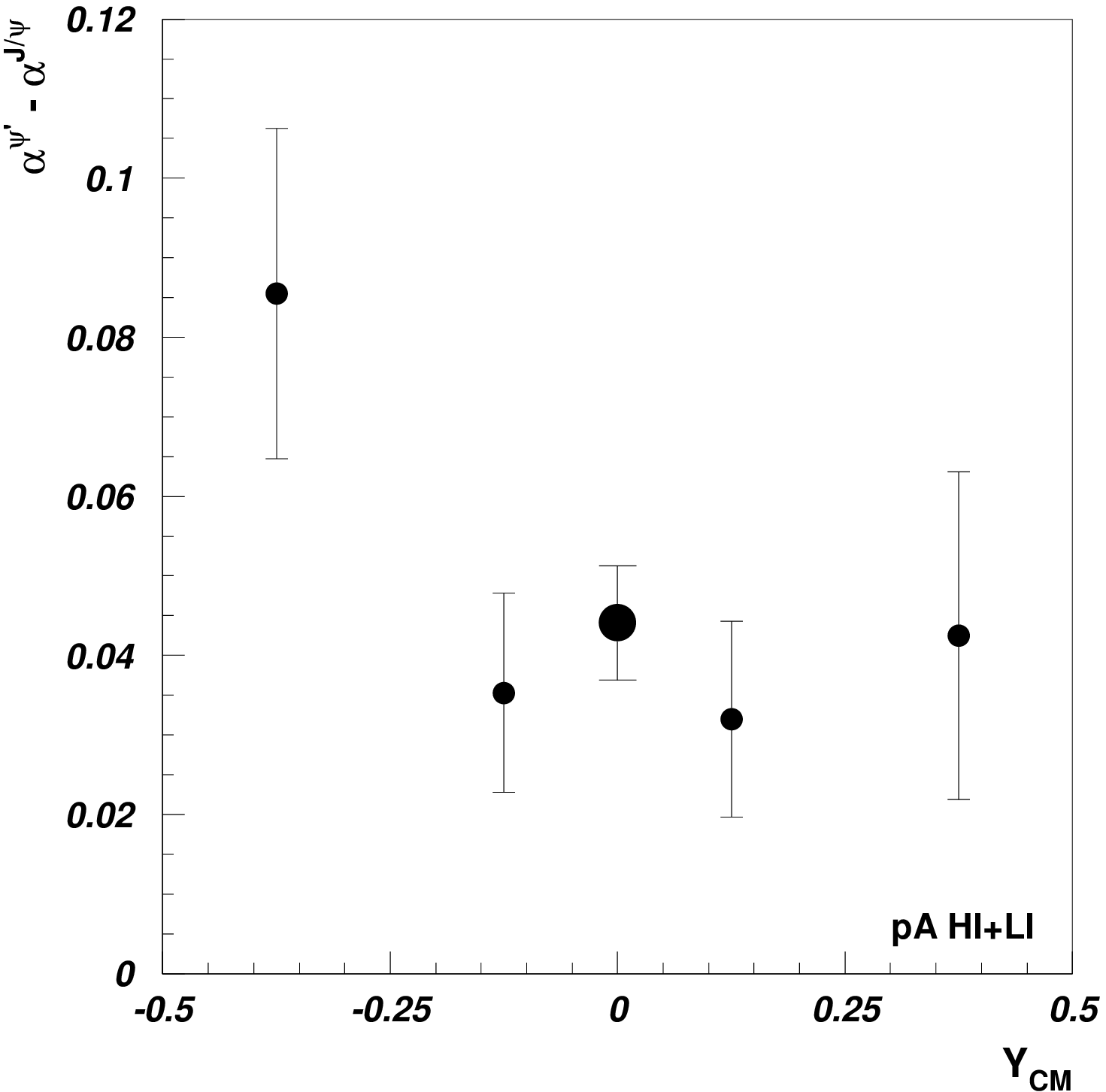}}
\resizebox{0.33\textwidth}{!}{%
\includegraphics*{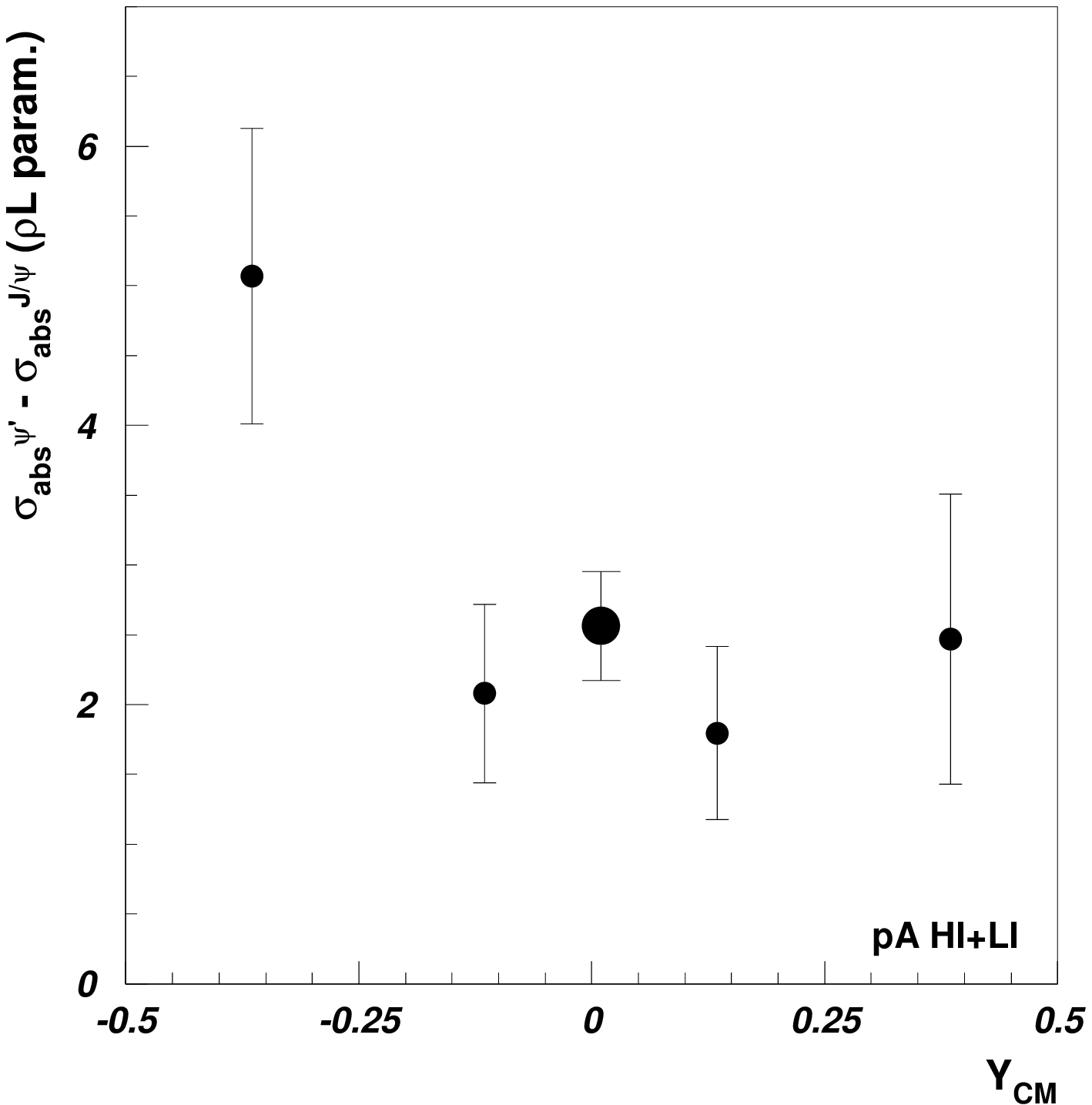}}
\end{tabular}
\vglue -2mm
\caption{Difference in the $\alpha$ parameters (Left) and in the absorption cross sections 
from the $e^{-<\rho L>\sigma_{abs}}$ parametrization for the $\jpsi$ and $\psip$ states as a function of
$Y_{CM}$. Obtained from the direct fit of the $\psip$ to $\jpsi$ cross sections ratio.
\label{y_drsgas_hili.eps}}
\end{figure}

Finally, Fig. \ref{y_drsgas_hili.eps} shows the difference between the $\alpha$ parameters and $\sigma_{abs}$
of the $\jpsi$ and $\psip$ states extracted from fitting directly the $\psip$ to $\jpsi$ cross section ratio (which cancels the
uncertainties in the normalization). The full Glauber model is not applicable to cross sections ratios, so we show only
the results of the $e^{-<\rho L>\sigma_{abs}}$ parametrization.

\section{Summary}
\label{sec:sum} 
The production cross sections of $\jpsi$ and $\psip$ in 450 GeV pA interactions was obtained and analyzed in the 
framework of charmonia nuclear absorption models. The $\jpsi$ nuclear absorption in the $-0.5<y_{cm}<0.5$ range is consistent with
previous NA50 results and yields $\alpha^{\jpsi} = 0.927\pm0.012$ (for $\sigma_{pA} = \sigma_{o} A^{\alpha}$ parametrization).
The fit with the full Glauber model yields $\sigma_{abs}^{\jpsi} = 4.7\pm0.8$ mb (or $4.3\pm0.7$ mb for the 
$e^{-<\rho L>\sigma_{abs}}$ parametrization). 
The corresponding fits for the $\psip$ show a stronger absorption than for the $\jpsi$, thus confirming the results
of the E866 collaboration:  $\alpha^{\psip}~=~0.886\pm0.014$ 
(with $\sigma_{abs}^{\psip} = 7.7\pm0.8$ and $6.7\pm0.8mb$ for the Glauber and the $e^{-<\rho L>\sigma_{abs}}$ 
models respectively).
The more precise fit of the ratio of $\psip$ to $\jpsi$ cross sections yields $\alpha^{\jpsi}-\alpha^{\psip} = 0.045\pm0.007$
and $\sigma_{abs}^{\psip} - \sigma_{abs}^{\jpsi} = 2.6\pm0.4~nb$ (for the $e^{-<\rho L>\sigma_{abs}}$ parametrization). No 
statistically significant dependence of the absorption on the rapidity is seen, although there is an indication of a stronger
absorption of the $\psip$ at small rapidities.

\section*{Acknowledgments}
The work was supported by the Funda\c{c}\~ao para a Ci\^encia e a Tecnologia
under the contract RPAXIS XXI/BD/16116/98. I appreciate the support of LIP
in attending this conference.

\end{document}